\documentclass{elsart}

\usepackage{amssymb}
\usepackage{cite}
\begin{document}

\begin{frontmatter}

% Title, authors and addresses

% use the thanksref command within \title, \author or \address for footnotes;
% use the corauthref command within \author for corresponding author footnotes;
% use the ead command for the email address,
% and the form \ead[url] for the home page:
% \title{Title\thanksref{label1}}
% \thanks[label1]{}
% \author{Name\corauthref{cor1}\thanksref{label2}}
% \ead{email address}
% \ead[url]{home page}
% \thanks[label2]{}
% \corauth[cor1]{}
% \address{Address\thanksref{label3}}
% \thanks[label3]{}

\title{On the Lagrangian and Hamiltonian description of the damped linear harmonic 
 oscillator}

\author{V. K. Chandrasekar, M. Senthilvelan, M. Lakshmanan}

\address{Centre for Nonlinear Dynamics, Department of Physics, 
Bharathidasan University, Tiruchirappalli - 620 024, India}

\date{\today} 

\begin{abstract}
Using the modified Prelle- Singer approach, we point out that explicit time
independent first integrals can be identified for the damped linear harmonic
oscillator in different parameter regimes. Using these constants of motion, an
appropriate Lagrangian and Hamiltonian formalism is developed and the resultant
canonical equations are shown to lead to the standard dynamical description.
Suitable canonical transformations to standard Hamiltonian forms are also 
obtained. It is also shown that a possible quantum mechanical description 
can be developed either in the coordinate or momentum representations  using the 
Hamiltonian forms.
 
\end{abstract}

\begin{keyword}
Lagrangian and Hamiltonian approach \sep Lagrangian and Hamiltonian mechanics 
\sep Canonical formalism, Lagrangians and variational principles 
\sep Quantum mechanics.
% keywords here, in the form: keyword \sep keyword

% PACS codes here, in the form: \PACS code \sep code
\PACS 11.10.Ef \sep 45.20.Jj \sep 04.20.Fy \sep 03.65.-w
\end{keyword}
\end{frontmatter}

\section{Introduction}
\label{1}
In a recent paper \cite{Chand1,Chand:2006}, we have studied the dynamical 
aspects of the Li\'enard type nonlinear oscillator equation 
\begin{eqnarray}
\ddot{x}+kx \dot{x}+\frac{k^2}{9}x^3+\lambda x=0,\label{emd}
\end{eqnarray}
where over dot denotes differentiation with respect to $t$ and $k$ and $\lambda$
are arbitrary parameters, and shown that the frequency of oscillations of a 
nonlinear oscillator need not depend upon the amplitude \cite{Chand1}. Our 
studies have also revealed that
the system (\ref{emd}), eventhough deceptively appears to be a dissipative one,
admits conservative Hamiltonian description for all values
of the parameters $k$ and $\lambda$. In
fact, the conventional criterion that the divergence of the flow function 
vanishes for a Hamiltonian system fails here in the original variables. These
facts motivate one to investigate further the possible existence of time 
independent Hamiltonian functions in actual dissipative systems. 

For our present study, we consider the ubiquitous damped linear harmonic
oscillator 
\begin{eqnarray}
\ddot{x}=-(\alpha \dot{x}+\lambda x)\equiv \phi(x,\dot{x}),\label{lam101}
\end{eqnarray} 
where $\alpha$ and $\lambda$ are the damping parameter and strength of the
forcing, respectively. When damping is absent, $\alpha=0$, the frequency of
oscillation $\omega=\sqrt{\lambda} >0$. For $\lambda <0$, one has unbounded
motion. System (\ref{lam101}) is known to possess a {\it time-dependent} Lagrangian
\cite{Bateman:1931,Ray:1979,Riewe:1996,Dekker:1981,Um:2002}, 
$L=e^{\alpha t}(\frac{1}{2}\dot{x}^2-\frac{\lambda}{2}x^2)$, 
and a Hamiltonian $H=\frac{1}{2}p^2 e^{-\alpha t}
+\frac{\lambda}{2}x^2e^{\alpha t}$. For the past several decades there has been
a number of attempts to quantize the damped linear harmonic oscillator (see for
example, Refs.
\cite{Dekker:1977,Dekker:1981,Gzyl:1983,Harris:1990,Srivastava:1991,Harris:1993
,Egger:1997,Karrlein:1997,Blasone:2001,Um:2002,Latimer:2002,Pereverzev:2003,
Choi:2003,Chruscinski:2006,Dito:2006})
from different 
points of view, but it appears that the
problem still eludes a completely satisfactory resolution. One of the major
conceptual difficulties seem to be the lack of an appropriate {\it 
time independent Hamiltonian formalism}. In this paper, we report such a
formalism by making use of the recently developed modified Prelle-Singer 
approach 
to identify integrals of motion and integrability of dynamical systems 
\cite{Cha:2004}.

System (\ref{lam101}) is obviously a dissipative one for $\alpha>0$ (and growing
one for $\alpha<0$), since
$\frac{d}{dt}(\frac{1}{2}\dot{x}^2 
+\frac{1}{2}\lambda x^2)=\frac{dE}{dt}=-\alpha \dot{x}^2<0$, where 
$E=(\frac{1}{2}\dot{x}^2 +\frac{1}{2}\lambda x^2)$ is
the total energy of the undamped oscillator. From a dynamical point of view Eq.
(\ref{lam101}) can be written as a system of coupled first order ordinary 
differential equations (ODEs)
%\begin{subequations}
\begin{eqnarray}
&&\dot{x}=y \label{emd1a}\\
&&\dot{y}=-\lambda x-\alpha y \label{emd1b}
\end{eqnarray}
%\label{emd1}
%\end{subequations}
so that the divergence of the flow function
\begin{eqnarray}
\Lambda=\frac{\partial \dot{x}}{\partial x}+\frac{\partial \dot{y}}{\partial y}
=-\alpha<0, \label{emd2}
\end{eqnarray}
which is the standard criterion for dissipative systems. Yet, we show here that
system (\ref{lam101}) possesses a time independent integral of motion for all
values of the parameters $\lambda$ and $\alpha$. We also provide the explicit 
form of the Lagrangian and Hamiltonian associated with Eq. (\ref{lam101}) in
different parametric regimes and derive all the expected solutions 
from the resulting canonical equations of motion. Our results also show that the
dissipative dynamical systems can also have conservative Hamiltonian
description, which can aid in appropriate quantization procedure as will be
indicated. 

The plan of the paper is as follows. In Section \ref{sec2}, we point out how the
time independent integrals can be obtained for the damped linear harmonic
oscillator (\ref{lam101}). Then in Section \ref{sec3}, we identify an appropriate
time-independent Lagrangian and so a  time-independent Hamiltonian in the
appropriate parameter ranges. In Section \ref{sec4}, we show that  through the
resulting canonical equations of motion, we obtain the correct explicit
solutions. In Section \ref{sec5}, we succeed to identify suitable canonical
transformations to  identify `standard' Hamiltonian description. Finally, in 
Section
\ref{sec6}, we discuss briefly the possible quantum formulation of the problem.
Section \ref{sec7} summarizes our results.

\section{Extended Prelle-Singer procedure}
\label{sec2}
%\subsection{PS method}
%\label{sec21}
To begin with we assume that the ODE (\ref{lam101}) 
admits a first integral $I(t,x,\dot{x})=C,$ with $C$ constant on the 
solutions, so that the total differential becomes
\begin{eqnarray}  
dI={I_t}{dt}+{I_{x}}{dx}+{I_{\dot{x}}{d\dot{x}}}=0, 
\label{met3}  
\end{eqnarray}
where each subscript denotes partial differentiation with respect 
to that variable. Rewriting Eq.~(\ref{lam101}) in the form 
$\phi dt-d\dot{x}=0$ and adding a null term 
$S(t,x,\dot{x})\dot{x}$ $ dt - S(t,x,\dot{x})dx$ to the latter, we obtain that on 
the solutions the 1-form
\begin{eqnarray}
(\phi +S\dot{x}) dt-Sdx-d\dot{x} = 0, \quad  \phi=-(\alpha \dot{x}+\lambda x).
\label{met6} 
\end{eqnarray}	
Hence, on the solutions, the 1-forms (\ref{met3}) and 
(\ref{met6}) must be proportional.  Multiplying (\ref{met6}) by the 
factor $ R(t,x,\dot{x})$ which acts as the integrating factors
for (\ref{met6}), we have on the solutions that 
\begin{eqnarray} 
dI=R(\phi+S\dot{x})dt-RSdx-Rd\dot{x}=0. 
\label{met7}
\end{eqnarray}
Comparing Eq.~(\ref{met3}) 
with (\ref{met7}) we have, on the solutions, the relations 
\begin{eqnarray} 
 I_{t}  = R(\phi+\dot{x}S),\quad 
 I_{x}  = -RS, \quad 
 I_{\dot{x}}  = -R.  
 \label{met8}
\end{eqnarray}
Then the compatibility conditions, 
$I_{tx}=I_{xt}$, $I_{t\dot{x}}=I_{{\dot{x}}t}$, $I_{x{\dot{x}}}=I_{{\dot{x}}x}$, 
between the Eqs.~(\ref{met8}), provide us \cite{Duarte}
\begin{eqnarray}          
S_t+\dot{x}S_x+\phi S_{\dot{x}} &=& 
   -\phi_x+\phi_{\dot{x}}S+S^2,\label {lin02}\\
R_t+\dot{x}R_x+\phi R_{\dot{x}} & =&
-(\phi_{\dot{x}}+S)R,\label {lin03}\\
R_x-SR_{\dot{x}}-RS_{\dot{x}}  &= &0.
\qquad \qquad\qquad \;\;\;\label {lin04}
\end{eqnarray}

Solving Eqs.~(\ref{lin02})-(\ref{lin04}) one can obtain expressions for $S$ and
$R$. It may be noted that any set of special solutions $(S,R)$ is sufficient for
our purpose. Once these forms are determined the integral of motion 
$I(t,x,\dot{x})$ can be deduced from the relation 
\begin{eqnarray}
 I= r_1
  -r_2 -\int \left[R+\frac{d}{d\dot{x}} \left(r_1-r_2\right)\right]d\dot{x},
  \label{met13}
\end{eqnarray}
where 
\begin{eqnarray} 
r_1 = \int R(\phi+\dot{x}S)dt,\quad
r_2 =\int (RS+\frac{d}{dx}r_1) dx. \nonumber
\end{eqnarray}
Equation~(\ref{met13}) can be derived straightforwardly by integrating the 
Eq.~(\ref{met8}).
%\subsection{The case $I_t=0$}
%\label{sec31}
%\subsubsection{Null forms}
%\label{sec311}

As our motivation is to explore time independent integral of motion for the 
Eq.~(\ref{lam101}) let us choose $I_t=0$.
In this case one can easily fix the null form $S$ from the first equation in
(\ref{met8}) as 
\begin{equation}
S = \frac{-\phi}{\dot{x}}=
\frac{(\alpha \dot{x}+\lambda x)}{\dot{x}}.
\label{mlin06}
\end{equation}
%\subsubsection{Integrating Factors}
%\label{sec312}
Substituting this form of $S$ into (\ref{lin03}) we get 
\begin{eqnarray}          
 \dot{x}R_x-(\alpha \dot{x}+\lambda x)R_{\dot{x}} 
 =-\frac{\lambda x}{\dot{x}}R.
 \label {lin03a}
\end{eqnarray}

Equation~(\ref{lin03a}) is a first order linear partial differential equation
with variable coefficients.
As we noted earlier any particular solution is
sufficient to construct an integral of motion (along with the function $S$).
To seek
a particular solution for $R$ one can make a suitable ansatz instead of looking
for the general solution. We assume $R$ to be of the form,
\begin{equation}
R = \frac{\dot{x}}{(A(x)+B(x)\dot{x}+C(x)\dot{x}^2)^r},
\label{mlin08}
\end{equation} 
where $A,\;B$ and $C$ are functions of their arguments, and $r$ is a constant
which are all to be
determined. {\it We demand the above form of  ansatz (\ref{mlin08}), which is very
important to derive the Hamiltonian structure associated with the given 
equation}, due to the following
reason. To deduce the first integral $I$ we assume a rational form for $I$, 
that is,
$I=\frac{f(x,\dot{x})}{g(x,\dot{x})}$, where $f$ and $g$
are arbitrary functions of $x$ and $\dot{x}$ and are independent of $t$, from
which we get
$I_x=\frac{f_{x}g-fg_x}{g^2}$ and $I_{\dot{x}}=\frac{f_{\dot{x}}g-fg_{\dot{x}}}{g^2}$.
From (\ref{met8}) one can see that $R=I_{\dot{x}}=\frac{f_{\dot{x}}g-fg_{\dot{x}}}{g^2},
\;S=\frac{I_x}{I_{\dot{x}}}=\frac{f_xg-fg_x}{f_{\dot{x}}g-fg_{\dot{x}}}$ and
$RS=I_x$, so that the denominator of the function $S$ should be the numerator of the
function $R$. Since the denominater of $S$ is $\dot{x}$ (vide Eq.~(\ref{mlin06}))
we fixed the numerator of $R$ as $\dot{x}$. To seek a suitable function in the 
denominator initially one can consider an
arbitrary form $R=\frac{\dot{x}}{h(x,\dot{x})}$. However, it is difficult to
proceed with this choice of $h$. So let us assume that $h(x,\dot{x})$ is a 
function which is polynomial in
$\dot{x}$. To begin with let us consider the case where $h$ is quadratic in 
$\dot{x}$,
that is, $h=A(x)+B(x)\dot{x}+C(x)\dot{x}^2$. Since $R$ is in rational form while
taking differentiation or integration the form of the denominator remains same 
but the  power of the denominator decreases or increases by a unit order from
that of the initial one. So instead of considering $h$ to be of the form
$h=A(x)+B(x)\dot{x}+C(x)\dot{x}^2$, one may consider a more general form
$h=(A(x)+B(x)\dot{x}+C(x)\dot{x}^2)^r$, where $r$ is a constant to be 
determined. The parameter $r$ plays an important role, as we see below. 
 
Substituting (\ref{mlin08}) into (\ref{lin03a}) and solving the resultant
equations, we arrive at the relation
\begin{eqnarray}
&&r\bigg[\dot{x}(A_x+B_x\dot{x}+C_x\dot{x}^2)-(\alpha \dot{x}+\lambda x)
(B+2C\dot{x})\bigg]\nonumber\\
&&\qquad\qquad\qquad\qquad\qquad\qquad=-\alpha(A+B\dot{x}+C\dot{x}^2).
\label{mlin09}
\end{eqnarray} 
Solving Eq.~(\ref{mlin09}), we can fix the forms of $A,\;B,\;C$ and $r$ and 
substituting them into Eq.~(\ref{mlin08}) we can get the integrating factor
$R$. Doing so, we find
\begin{eqnarray}
R=\left\{
\begin{array}{ll}
\displaystyle{\frac{\dot{x}}{(\lambda x^2+\alpha x\dot{x}+\dot{x}^2)}},
&\;\;\alpha^2 <4\lambda \\
\displaystyle{\frac{\dot{x}}{(\dot{x}+\frac{(r-1)}{r}\alpha x)^{r}}},
&\;\;\alpha^2\geq4\lambda  \\
\displaystyle{\dot{x}}, & \;\; \alpha=0 
\end{array}\right.\label{lam107}
\end{eqnarray} 
where 
\begin{eqnarray}
r=\frac{\alpha}{2\lambda}\bigg[\alpha\pm\sqrt{\alpha^2-4\lambda}\bigg].
\label{lam107a}
\end{eqnarray} 
One can easily check the functions $S$ and $R$ given in (\ref{mlin06}) and 
(\ref{lam107}), respectively, satisfy (\ref{lin04}) also. Finally, substituting 
$R$ and $S$ into the form (\ref{met13}) for the integral we get 
\begin{eqnarray}
I=\left\{
\begin{array}{ll}
\frac{1}{2}\log(\dot{x}^2+\alpha x\dot{x}+\lambda x^2)+\frac{\alpha }
{2\omega}tan^{-1}\bigg[\frac{\alpha \dot{x}+2\lambda x}{2\omega
\dot{x}}\bigg],\\
\qquad \qquad \qquad 
\alpha^2 <4\lambda \quad (\lambda,\alpha\neq0) \quad \mbox{(underdamped)} \\
\frac{(r-1)}{(r-2)}(\dot{x}+\frac{\alpha}{r}x)
(\dot{x}+\frac{(r-1)}{r}\alpha x)^{(1-r)},\\
\qquad \qquad \qquad
\alpha^2>4\lambda \quad (\lambda,\alpha\neq0) \quad \mbox{(overdamped)}\\
\frac{\dot{x}}{(\dot{x}+\frac{1}{2}\alpha x)}
-\log(\dot{x}+\frac{1}{2}\alpha x), \\
\qquad \qquad \qquad 
\alpha^2=4\lambda \quad (\lambda,\alpha\neq0) \quad \mbox{(critically damped)} \\
\dot{x}+\alpha x, \\
\qquad \qquad \qquad
\lambda=0 \quad (\alpha\neq0) \quad \mbox{(pure damping)} \\
\dot{x}^2+\lambda x^2,\\
\qquad \qquad \qquad 
\alpha=0 \quad (\lambda\neq0) \quad \mbox{(no damping)}
\end{array}\right.\label{mlin12}
\end{eqnarray}
where $\omega=\frac{1}{2}\sqrt{(4\lambda-\alpha^2)}$. One can easily check
(using Eq. (\ref{lam101})) that $\frac{dI}{dt}=0$ for each of the cases in 
Eq.~(\ref{mlin12}). In the above, from a physical point of view we consider
only the principle branches to avoid multivaluedness. 

Eq.~(\ref{mlin12}) demonstrates clearly that the damped linear harmonic
oscillator (\ref{lam101}) admits a time independent integral of motion for all
values of the parameters $\alpha$ and $\lambda$.

Now, due to the fact that the integrals of motion (\ref{mlin12}) are time independent
ones we can look for a Hamiltonian description for the respective equations of
motion. In fact, in the following we do obtain the explicit time independent 
Hamiltonians for all the above cases.
\section{Hamiltonian description of the damped linear harmonic oscillator}
\label{sec3}
Assuming the existence of a Hamiltonian
\begin{eqnarray}
I(x,\dot{x}) = H(x,p) = p \dot{x}-L(x,\dot{x}),
\label{mlin13}
\end{eqnarray}
where $L(x,\dot{x})$ is the Lagrangian and $p$ is the canonically conjugate
momentum, we have
\begin{eqnarray}
&&\frac{\partial{I}}{\partial\dot{x}} = \frac{\partial{H}}{\partial\dot{x}} = 
\frac{\partial{p}}{\partial\dot{x}}\dot{x} + p - \frac{\partial{L}}{\partial\dot{x}} = 
\frac{\partial{p}}{\partial\dot{x}} \dot{x}.
\label{mlin13a}
\end{eqnarray}
From (\ref{mlin13a}) we identify
\begin{eqnarray}
p =\int \frac{I_{\dot{x}}}{\dot{x}} d\dot{x}+f(x)
\label{mlin13b}
\end{eqnarray}
where $f(x)$ is an arbitrary function of $x$. Equation (\ref{mlin13b}) has 
also been derived recently by a different methodology 
\cite{Lopez1}. Hereafter, without loss of generality we take $f(x)=0$.
Substituting the known expression of $I$ into (\ref{mlin13b}) and integrating 
it we can obtain the expression for the canonical momentum $p$. Substituting 
back the latter into the Eq.~(\ref{mlin13}) and simplifying the resultant 
equation we arrive at the following Lagrangian 
\begin{eqnarray}
L=\left\{
\begin{array}{ll}
\frac{1}{2\omega}
\bigg(tan^{-1}\bigg[\frac{2\dot{x}+\alpha x}{2\omega x}\bigg]
(\frac{2\dot{x}}{x})+\alpha tan^{-1}\bigg[\frac{\alpha \dot{x}+2\lambda x}
{2\omega \dot{x}}\bigg]\bigg)\\
\qquad \qquad -\frac{1}{2}\log(\dot{x}^2+\alpha x\dot{x}+\lambda x^2),
&\;\;\alpha^2 <4\lambda \\
\frac{1}{(2-r)}(\dot{x}+\frac{(r-1)}{r}\alpha x)^{(2-r)},
&\;\;\alpha^2>4\lambda \;\;(r\neq0,1,2),\\
\log(\dot{x}+\frac{1}{2}\alpha x), & \;\; \alpha^2=4\lambda\;\; (r=2) \\
\dot{x}\log(\dot{x})-\dot{x}-\alpha x, &\;\;\lambda=0\;\; (r=1)\\
\dot{x}^2-\lambda x^2,&\;\; \alpha=0\;\;(r=0)
\end{array}\right.\label{mlin14}
\end{eqnarray}
and Hamiltonian
\begin{eqnarray}
H=\left\{
\begin{array}{ll}
\frac{1}{2}\log[x^2 sec^2(\omega xp )]-\frac{\alpha }{2}xp,
&\;\;\alpha^2 <4\lambda \\
\frac{(r-1)}{(r-2)}(p)^{\frac{(r-2)}{(r-1)}}-\frac{(r-1)}{r}\alpha xp,
&\;\;\alpha^2>4\lambda \;\;(r\neq0,1,2),\\
\log(p)-\frac{1}{2}\alpha xp, 
& \;\; \alpha^2=4\lambda\;\; (r=2) \\
e^{p}+\alpha x,
&\;\;\lambda=0\;\; (r=1)\\
\frac{1}{2}p^2+ \frac{\lambda}{2}x^2,
&\;\; \alpha=0\;\;(r=0)
\end{array}\right.\label{mlin16}
\end{eqnarray}
where $p$ is the canonically conjugate momentum 
\begin{eqnarray}
p=\left\{
\begin{array}{ll}
\frac{tan^{-1}\bigg[\frac{2\dot{x}+\alpha x}{2\omega x}\bigg]}
{\omega x},&\;\;\alpha^2 <4\lambda \\
(\dot{x}+\frac{(r-1)}{r}\alpha x)^{(1-r)},
&\;\;\alpha^2\geq4\lambda \;\;(r\neq0,1),\\
\log(\dot{x}), &\;\;\lambda=0\;\; (r=1)\\
p=\dot{x},&\;\; \alpha=0\;\;(r=0)
\end{array}\right.\label{mlin15}
\end{eqnarray}
respectively, for the damped harmonic oscillator (\ref{lam101}).

One can easily check that the canonical equations of motion for the above
Hamiltonians are nothing but the equation of motion (\ref{lam101}) of the 
damped linear harmonic oscillator in the appropriate parametric regimes 
(see below for details). 
\section{Canonical Equations and Solutions}
\label{sec4}
We now write down the canonical equations of motion obtained from the above
Hamiltonian expressions for each of the cases separately and obtain the 
corresponding solutions, as it should be, to show the correctness of our
results.
\subsection{Case: 1 $\alpha^2 <4\lambda$:}
Now from the form of the Hamiltonian given in 
Eq. (\ref{mlin16}), we first derive the Hamilton's equation for the case 
$\alpha^2 <4\lambda$:
\begin{eqnarray}
&&\dot{x}=\frac{1}{2}x(2\omega \tan(\omega xp)-\alpha),\label{mlin17b}\\
&&\dot{p}
=\frac{1}{2}p(\alpha-2\omega \tan(\omega xp))-\frac{1}{x},
\label{mlin17a}
\end{eqnarray}
where $\omega=\frac{1}{2}\sqrt{(4\lambda-\alpha^2)}$. Rewriting 
Eq. (\ref{mlin17b}) for $p$ and substituting it into Eq. (\ref{mlin17a}), one can 
immediately check that the equation of motion
(\ref{lam101}) indeed follows straightforwordly from the canonical Eqs.~ 
(\ref{mlin17b})-(\ref{mlin17a}). Now multiplying Eq. (\ref{mlin17a}) by $x$ 
and (\ref{mlin17b}) by $p$ and adding and integrating the resultant equation 
we get
\begin{eqnarray}
p=\frac{\delta-t}{x},\label{mlin17c}
\end{eqnarray}
where $\delta$ is an integration constant. Substituting the expression for $p$ 
into 
Eq.~(\ref{mlin17b}) we obtain the following equation for $\dot{x}$, namely, 
\begin{eqnarray}
\dot{x}=\frac{1}{2}x(2\omega \tan(\omega (\delta-t))-\alpha),
\label{mlin17d}
\end{eqnarray}
which upon integration leads us to the general solution for the Eq.~
(\ref{lam101}), for the choice $\alpha^2<4\lambda$, of the form 
\begin{eqnarray}
x(t)=C_0e^{-\frac{1}{2}\alpha t}\cos(\omega (t-\delta)),
\label{mlin17e}
\end{eqnarray}
where $C_0\;(>0)$ is the second integration constant, as it should be for the
under damped case. We also note that on substitution of the solution 
(\ref{mlin17e}) into (\ref{mlin12}), one gets 
$I=H=(\frac{1}{2}\log{C_0^2}-\frac{1}{2}\alpha\delta)$.
\subsection{Case: 2  $\alpha^2 >4\lambda$ ($r\neq0,1,2$ cases):}
In this case we have the following Hamilton's equation of motion
\begin{eqnarray}
&&\dot{x}=(p)^{\frac{1}{(1-r)}}
-\frac{(r-1)}{r}\alpha x.\label{mlin18b}\\
&&\dot{p}=\frac{(r-1)}{r}\alpha p\label{mlin18a}
\end{eqnarray}
Again it is quite straightforward to check that Eqs.~ (\ref{mlin18b}) 
and (\ref{mlin18a}) are equivalent to (\ref{lam101}) for 
$\alpha^2 >4\lambda$ ($r\neq0,1,2$).
 
Integrating the above equations, we obtain the general 
solution as
\begin{eqnarray}
x(t)=C_0e^{-\frac{(r-1)}{r}\alpha t}+C_1\frac{r}{(r-2)\alpha }
e^{-\frac{1}{r}\alpha t},\label{mlin18c}
\end{eqnarray}
where $C_0$ and $C_1\;(>0)$ are integration constants. Eq. (\ref{mlin18c}) is the well
known solution for the overdamped case. Again, we check that from 
(\ref{mlin12}) that $I=H=\bigg(\frac{1-r}{r}\bigg)\alpha C_0C_1^{(1-r)}$. 
\subsection{Case: 3 $\alpha^2 =4\lambda$ ($r=2$):}
Here the Hamilton's equations are
\begin{eqnarray}
\dot{x}=\frac{1}{p}-\frac{\alpha}{2} x, \qquad 
\dot{p}=\frac{\alpha}{2} p.
\end{eqnarray}
The general solution in this case is 
\begin{eqnarray}
x(t)=(C_0+C_1t)e^{-\frac{1}{2}\alpha t},\label{mlin19}
\end{eqnarray}
where $C_0$ and $C_1\;(>0)$ are integration constants. Here 
$I=H-1=1-\frac{1}{2}\alpha C_0C_1^{-1}-\log{C_1}$.
\subsection{Case: 4 $\alpha\neq0,\;\lambda=0$ ($r=1$):}
From (\ref{mlin16}), the canonical equations for this case are
\begin{eqnarray}
\dot{x}=e^{p}, \qquad \dot{p}=-\alpha,
\end{eqnarray}
leading to the correct solution for this case,
\begin{eqnarray}
x(t)=C_0-\frac{1}{\alpha}e^{-\alpha(t-t_0)},\label{mlin20}
\end{eqnarray}
where $C_0$ and $t_0$ are integration constants. Here 
$I=H=\alpha C_0$.

Also it may be that in each of the four cases for the associated canonical
equations, the divergence of the flow function 
\begin{eqnarray}
\Lambda=\frac{\partial \dot{x}}{\partial x}
+\frac{\partial \dot{p}}{\partial p}=0,\label{div01}
\end{eqnarray}
as it should be in view of the existence of conserved Hamiltonian function.
\section{Standard Hamiltonian form}
\label{sec5}
The Hamiltonian forms given in Eq. (\ref{mlin16}) for the various parametric 
ranges appear
rather complicated: however, they can be brought to `standard' forms by effecting
appropriate canonical transformations.
\subsection{Case: 1 $\alpha^2 <4\lambda$:}
With the canonical transformation $P=\frac{1}{2}\log(x^2)$ and $U=-xp$, the 
Hamiltonian for the
present case, $H=\frac{1}{2}\log[x^2 sec^2(\omega xp )]-\frac{\alpha }{2}xp$, 
can be transformed into
\begin{eqnarray}
H=P+\frac{1}{2}\log[sec^2(\omega U)]+\frac{\alpha}{2}U.\label{sh01}
\end{eqnarray}
\subsection{Case: 2  ($\alpha^2 >4\lambda$) and Case: 3 ($\alpha^2=4\lambda$):}
For these two cases, the Hamiltonian functions given in Eq. (\ref{mlin16}) get
transformed through the canonical transformation $x=-Pe^{-U}$ and $p=e^{U}$ as 
\begin{eqnarray}
H=\left\{
\begin{array}{ll}
P+\frac{r}{(r-2)\alpha}e^{\frac{(r-2)}{(r-1)}U},
&\;\;\alpha^2>4\lambda \;\;(r\neq0,1,2),\\
P+\frac{2}{\alpha} U, 
& \;\; \alpha^2=4\lambda.
\end{array}\right.\label{sh02}
\end{eqnarray}
\subsection{Case: 4 $\alpha\neq0,\;\lambda=0$ ($r=1$):}
In this case, the Hamiltonian $H=e^{p}+\alpha x$ transformed into the form
\begin{eqnarray}
H=P+e^{-\alpha U},\label{sh03}
\end{eqnarray}
through the canonical transformation $x=\frac{P}{\alpha}$ and $p=-\alpha U$.

Since the canonical variables are now separated, the new Hamiltonians can be 
utilized for applications including the quantization.
 
Before proceeding further, we wish to point out that an explicit
time independent integral for the damped harmonic oscillator (\ref{lam101}) 
was given
by L\'opez and L\'opez \cite{Lopez1,Lopez2}. However, no explicit Hamiltonian could
be associated with it. We trace the reason for this in Appendix-A.

\section{Quantum description}
\label{sec6}
Having obtained appropriate Hamiltonian formalism, we now proceed to identify a
possible quantum description of the damped linear harmonic oscillator. For this
purpose, we first rewrite the classical Hamiltonian in a symmetrized form so as
to ensure hermiticity when the dynamical variables are represented by linear
operators acting on a quantum Hilbert space. Thus we write
\begin{eqnarray}
H=\left\{
\begin{array}{ll}
\frac{1}{2}\log[x^2 sec^2( \frac{\omega}{2}(xp+px) )]-\frac{\alpha }{4}(xp+px),
&\;\;\alpha^2 <4\lambda \\
\frac{(r-1)}{(r-2)}(p)^{\frac{(r-2)}{(r-1)}}-\frac{(r-1)}{2r}\alpha (xp+px),
&\;\;\alpha^2>4\lambda \;\;(r\neq0,1,2),\\
\log(p)-\frac{\alpha}{4}(xp+px), 
& \;\; \alpha^2=4\lambda\;\; (r=2) \\
e^{p}+\alpha x,
&\;\;\lambda=0\;\; (r=1)\\
\frac{1}{2}p^2+ \frac{\lambda}{2}x^2,
&\;\; \alpha=0\;\;(r=0).
\end{array}\right.\label{tap01}
\end{eqnarray}
In the above $x$ and $p$ are now considered as linear operators satisfying the
cannonical commutation rule (CR)
\begin{eqnarray}
[x,p]=i\hslash.\label{tsc01b}
\end{eqnarray}

Obviously, one can look for a coordinate representation in which $x$ is treated
as a variable and $p$ as the operator $-i\hslash\frac{\partial }{\partial x}$ so
that the CR (\ref{tsc01b}) is satisfied or in a momentum representation one may
treat $p$ as a c-number and $x$ as the operator 
$x=i\hslash\frac{\partial }{\partial p}$ satisfying (\ref{tsc01b}). For the
present problem we choose a momentum representation and write the time-dependent
Schr$\ddot{o}$dinger equation as
\begin{eqnarray}
i\hslash\frac{\partial \Psi(p,t)}{\partial t}=H\Psi(p,t).\label{tsc01}
\end{eqnarray}
Next we look for the stationary states
\begin{eqnarray}
\Psi(p,t)=\psi(p)e^{\displaystyle -\frac{i\mathcal{E}t}{\hslash}}\label{tsc02}
\end{eqnarray}
so that (\ref{tsc01}) reduces to the time-independent Schr$\ddot{o}$dinger 
equation in the momentum space:
\begin{eqnarray}
H\psi(p)=\mathcal{E}\psi(p).
\label{tsc03}
\end{eqnarray}
Here $\mathcal{E}$ is identified as the eigenvalue corresponding to the
Hamiltonian operator $H$. 
In the following we first consider the pure damped case, that is, $\lambda=0$ 
and then the remaining cases (except the case $\alpha^2 <4\lambda$).
We wish to note one can consider the quantization of the transformed 
Hamiltonians (\ref{sh01})-(\ref{sh03}) also which is discussed in Appendix B.

\subsection{Case: 1 $\lambda=0$:}
In the case $\lambda=0$, the Hamiltonian $H=e^{p}+\alpha x$ is quantized 
so that the time independent Schr$\ddot{o}$dinger equation in
the momentum space becomes
\begin{eqnarray}
H\psi=\mathcal{E}\psi
\Rightarrow (i\alpha\hslash \frac{d}{dp}+e^{p})\psi=\mathcal{E}\psi.
\label{mlin21}
\end{eqnarray}
Solving the above spectral problem for any real value of $\mathcal{E}$, $-\infty
<\mathcal{E}<\infty$, we obtain the wavefunction in the form
\begin{eqnarray}
\psi(p)=C_1\exp \left[\displaystyle i\bigg(e^{p}-\frac{p\mathcal{E}
}{\alpha\hslash}\bigg)\right],
\label{1mlin24b}
\end{eqnarray}
where $C_1$ is the normalization constant. From equations (\ref{tsc02}) and 
(\ref{1mlin24b}) we get the plane wave like solution 
\begin{eqnarray}
\Psi(p,t)=C_1\exp \left[\displaystyle -\frac{i\mathcal{E}t}{\hslash}
+i\bigg(\frac{e^{p}-p\mathcal{E}}{\alpha\hslash}\bigg)\right], \quad -\infty
<\mathcal{E}<\infty
\label{2mlin24b}
\end{eqnarray}
as the final solution of (\ref{tsc03}).

\subsection{Case: 2 $\alpha^2 =4\lambda$:}
For the case $\alpha^2=4\lambda$, the eigen value equation becomes 
\begin{eqnarray}
H\psi=\mathcal{E}\psi \Rightarrow \bigg(\log(p)-i\frac{\hslash \alpha}{4}(1
+2p\frac{d }{d p})\bigg)\psi=\mathcal{E}\psi.
\label{tap02}
\end{eqnarray}
Introducing the transformation of the form 
$\hat{\psi}=\psi e^{i\frac{1}{\hslash \alpha}(\log(p))^2}$ into the above
equation we get 
\begin{eqnarray}
p\frac{d\hat{\psi}}{d p}
=\bigg[-\frac{1}{2}+i\frac{2 \mathcal{E}}{\hslash\alpha}\bigg]\hat{\psi},
\quad -\infty <\mathcal{E}<\infty.
\label{tap02a}
\end{eqnarray}
The above eigenvalue equation is exactly of the same form as that of the eigenvalue
equation in the {\it coordinate space} for the toy model for quantum damping
introduced by Chruscinski \cite{Chruscinski:2003}, see Eq. (6.2), in a rigged
Hilbert space.

Consequently the generalized eigenfunctions can be given as 
\begin{eqnarray}
\hat{\psi}(p):=\frac{1}{\sqrt{-\pi\alpha\hslash}}
p_{\pm}^{\displaystyle i\frac{2\mathcal{E}}{\alpha\hslash}-\frac{1}{2}},
\label{tap03}
\end{eqnarray}
where following Ref. \cite{Chruscinski:2003} the tempered-distributions are
defined as
\begin{eqnarray}
p_{+}^{\lambda}:=\left\{
\begin{array}{ll}
p^{\lambda},&\;\;p\geq0, \\
0,&\;\;p<0,
\end{array}\right.\label{tap01aa}
\end{eqnarray}
\begin{eqnarray}
p_{-}^{\lambda}:=\left\{
\begin{array}{ll}
0,&\;\;p\geq0, \\
|p|^{\lambda},&\;\;p<0,
\end{array}\right.\label{tap01bb}
\end{eqnarray}
with $\lambda \in \mathbf C$. The resultant wavefunction becomes
\begin{eqnarray}
\psi(p)=\frac{1}{\sqrt{-\pi\alpha\hslash}}
p_{\pm}^{\displaystyle \bigg(i\frac{2\mathcal{E}}{\alpha\hslash}
-\frac{1}{2}\bigg)}
e^{\displaystyle -i\bigg(\frac{(\log(p))^2}{\hslash \alpha}\bigg)}.
\label{tap04}
\end{eqnarray}

In Ref. \cite{Chruscinski:2003}, Chruscinski has shown that the complex 
eigenvalues of the underlying Hamiltonian
correspond to the poles of energy eigenvectors when continued to
the complex eigenvalue plane and that the corresponding generalized
eigenvectors may be interpreted as resonant states and these resonant states
are responsible for the irreversible quantum dynamics. Consequently, similar
arguments can hold good here also. Detailed description will be given elsewhere.

\subsection{Case: 3 $\alpha^2 >4\lambda$:}
In this case we have the following eigen value equation to be solved, 
\begin{eqnarray}
H\psi=\mathcal{E}\psi \Rightarrow \bigg(\frac{(r-1)}{(r-2)}(p)^{\frac{(r-2)}{(r-1)}}
-i\frac{(r-1)}{2r}\alpha\hslash (1+2p\frac{d }{d p})\bigg)\psi
=\mathcal{E}\psi.
\label{tap08}
\end{eqnarray}
As we did earlier let us introduce the transformation
\begin{eqnarray} 
\hat{\psi}=\psi e^{\displaystyle i
\bigg(\frac{r(r-1)}{\alpha\hslash(r-2)^2}p^{\frac{r-2}{r-1}}
\bigg)},\label{tap08b}
\end{eqnarray} 
into equation (\ref{tap08}) and we obtain the eigenvalue equation 
\begin{eqnarray}
p\frac{d\hat{\psi}}{d p}
=-\bigg[i\frac{r\mathcal{E}}{(1-r)\alpha\hslash}+\frac{1}{2}\bigg]\hat{\psi},
\quad -\infty <\mathcal{E}<\infty.
\label{tap08a}
\end{eqnarray}
Correspondingly, again following Chruscinski \cite{Chruscinski:2003} 
the generalized eigenfunctions can be given as 
\begin{eqnarray}
\hat{\psi}(p):=\sqrt{\frac{r}{2(1-r)\pi\alpha\hslash}}\;
p_{\pm}^{\displaystyle-\bigg(i\frac{r\mathcal{E}}{(1-r)\alpha\hslash}
+\frac{1}{2}\bigg)}.
\label{tap09}
\end{eqnarray}
Then the wavefunction becomes
\begin{eqnarray}
\psi(p)=\sqrt{\frac{r}{2(1-r)\pi\alpha\hslash}}\;
p_{\pm}^{\displaystyle -\bigg(i\frac{r\mathcal{E}}{(1-r)\alpha\hslash}
+\frac{1}{2}\bigg)}e^{\displaystyle -i
\bigg(\frac{r(r-1)}{\alpha\hslash(r-2)^2}p^{\frac{r-2}{r-1}}
\bigg)}.
\label{tap10}
\end{eqnarray}
Again the properties of the generalized eigenfunctions discussed in Ref. 
\cite{Chruscinski:2003} hold good here also.

Finally for the case $\alpha^2 <4\lambda$ in Eq. (\ref{tap01}), it is not yet 
clear how to solve the linear eigenvalue problem. However, another possible
approach is to make use of the canonically transformed Hamiltonian of the form
(\ref{sh01}) as indicated in Appendix B.

\section{Conclusions}
\label{sec7}
In this paper, we have identified explicit time independent first integrals 
for the 
damped linear harmonic oscillator valid in different parameter regimes using 
the modified Prelle- Singer approach. We have constructed the 
appropriate Lagrangians and explicit Hamiltonians from the time independent first 
integrals and transformed the corresponding Hamiltonian forms to standard 
Hamiltonian forms using suitable canonical transformations.
We have also indicated that a quantized description can be developed  
in the momentum representation using the Hamiltonian forms 
and map onto known quantum mechanical toy models of damped systems. We expect
that further analysis of the quantum case can reveal much information on the
quantum damped oscillator. 

\section*{Acknowledgments}
The authors wish to thank Professor R. Simon, Institute of Mathematical
Sciences, Chennai for a useful discussion.
The work of VKC is supported by CSIR in the form of a CSIR Senior Research
Fellowship.  The work of ML forms part of a Department of Science and 
Technology, Government of India sponsored research project and is supported by a
Department of Atomic Energy Raja Ramanna Fellowship.

\appendix

\section{Appendix}
\label{sec8}
A set of first integrals for the damped harmonic oscillator (\ref{lam101}) was
reported in Refs. \cite{Lopez1,Lopez2} in the form
%\begin{mathletters}
\begin{eqnarray}
K=\frac{1}{2}(\dot{x}^2+\alpha x\dot{x}+\lambda x^2)e^{-\alpha G}.
\label{lop01a}
\end{eqnarray} 
where
\begin{eqnarray}
G=\left\{
\begin{array}{ll}
\frac{2}{\sqrt{4\lambda-\alpha^2}}
tan^{-1}\bigg[\frac{2\dot{x}+\alpha x}{\sqrt{4\lambda-\alpha^2} x}\bigg]
,&\;\;\alpha^2 <4\lambda\;\mbox{(underdamped)}\\
\frac{1}{\sqrt{\alpha^2-4\lambda}}
\log\bigg(\frac{(\alpha-\sqrt{\alpha^2-4\lambda})x+2\dot{x}}
{(\alpha+\sqrt{\alpha^2-4\lambda})x+2\dot{x}}\bigg),
&\;\;\alpha^2>4\lambda \;\mbox{(overdamped)}\\
\frac{4 \dot{x}}{2\alpha\dot{x}+\alpha^2 x}, 
& \;\; \alpha^2=4\lambda\;\mbox{(critically damped)} 
\end{array}\right.\label{lop01b}
\end{eqnarray}
%\label{lop01}
%\end{mathletters}
Note that in Refs. \cite{Lopez1,Lopez2} the value of $G$  for the critically 
damped case is given as $G=\displaystyle{\frac{1}{\omega_{\alpha}+\frac{v}{x}}}$
but the correct form is
$G=\displaystyle{\frac{1}{\omega_{\alpha}\frac{x}{v}(\omega_{\alpha}
+\frac{v}{x})}}$.
  
All the first integrals given in Eq.~(\ref{lop01a}) are related to our 
first integrals (Eq.~(\ref{mlin12})-underdamped, overdamped, critically damped 
cases) in the following manner (after using suitable transformation formulas)
\begin{eqnarray}
I=\left\{
\begin{array}{ll}
\frac{1}{2}\log(2K)-\frac{\alpha }
{\sqrt{4\lambda-\alpha^2}}tan^{-1}[\frac{\sqrt{4\lambda-\alpha^2}}{\alpha}]
,&\;\;\alpha^2 <4\lambda\;\mbox{(underdamped)} \\
\frac{(r-2)}{(r-1)}(2K)^{\frac{(2-r)}{2}},
&\;\;\alpha^2>4\lambda \;\mbox{(overdamped)}\\
-\frac{1}{2}\log(2K), 
& \;\; \alpha^2=4\lambda\;\mbox{(critically damped)} 
\end{array}\right.\label{mlin12a}
\end{eqnarray}
where $r$ is the quantity defined in Eq.~(\ref{lam107a}).

In Refs. \cite{Lopez1,Lopez2}, the authors have found the Lagrangian and an
implicit Hamiltonian only for underdamped case ($\alpha^2 <4\lambda$) and 
concluded that {\it the dissipative harmonic oscillator has nonexplicit 
Hamiltonian formulation}. For the other cases it was stated that {\it it is not 
possible to get a close expression for the Lagrangian}. However in the present 
work we have found the Lagrangian and explicit Hamiltonian for all the 
three cases.

The reason for not getting the Lagrangian and generalized linear momentum from
the first integral (\ref{lop01a}) reported in Refs. \cite{Lopez1,Lopez2} 
is as follows:

For example, the first integral $I$ given in Eq. (\ref{mlin12}) for the 
overdamped case is related to the first integral $K$ reported in  Refs.
\cite{Lopez1,Lopez2} through the relation 
$I=\frac{(r-2)}{(r-1)}(2K)^{\frac{(2-r)}{2}}$. Our first integral $I$  for the
overdamped case unambiguously leads to the generalized linear momentum, 
Lagrangian and explicit Hamiltonian from Eqs.~ (\ref{mlin13}) and  
(\ref{mlin13b}) whereas integral  $K$ does not. The reason for the above
is not difficult to understand. For example, let us consider the first
integral  $I=\dot{x}^2+\omega^2 x^2$ for the linear harmonic oscillator which
gives the conjugate momentum $p=\dot{x}$ and Lagrangian 
$L=\dot{x}^2-\omega^2 x^2$ (vide Eqs. (\ref{mlin13}) and (\ref{mlin13b})).
However, $\hat{I}=(\dot{x}^2+\omega^2 x^2)^n$, where $n$ is some arbitrary 
parameter,  is also a first integral for SHO.  By substituting this first
integral  $\hat{I}$ into the expressions (\ref{mlin13}) and (\ref{mlin13b}),
the  linear momentum and Lagrangian cannot be obtained in terms of any closed
form.  In order to find the Lagrangian and Hamiltonian from the known first
integral, one  has to take appropriate form of the first integral. Our PS
method gives directly the  correct form of first integral needed to find the 
Lagrangian and Hamiltonian of the system.

\section{Quantum description in standard Hamiltonian form}
\label{sec9}
%\subsection{Case: 1 :}
Let us consider the Eqs. (\ref{sh01}) - (\ref{sh03}), which are all of the 
form
\begin{eqnarray}
H=P+g(U),\label{qsh01}
\end{eqnarray}
where
\begin{eqnarray}
g(U)=\left\{
\begin{array}{ll}
\frac{1}{2}\log[sec^2(\omega U)]+\frac{\alpha}{2}U,&\;\;\alpha^2>4\lambda \\
\frac{r}{(r-2)\alpha}e^{\frac{(r-2)}{(r-1)}U},
&\;\;\alpha^2>4\lambda \;\;(r\neq0,1,2),\\
\frac{2}{\alpha} U, 
& \;\; \alpha^2=4\lambda\\
e^{-\alpha U}, & \lambda=0.
\end{array}\right.\label{qsh02}
\end{eqnarray}
The Hamiltonian (\ref{qsh01}) is now quantized in the new 
coordinate space, so that here $P=-i\hslash \frac{\partial }{\partial U}$ and
the corresponding time independent Schr$\dot{o}$dinger equation becomes
\begin{eqnarray}
H\psi=E\psi
\Rightarrow (-i\hslash \frac{d }{d U}+g(U))\psi=\mathcal{E}\psi.
\label{qsh03}
\end{eqnarray}
Introducing the transformation 
$\hat{\psi}=\psi e^{i\frac{\int g(U)dU}{\hslash }}$ into (\ref{qsh03}) we get 
\begin{eqnarray}
\frac{d\hat{\psi}}{d U}
=i\frac{\mathcal{E}}{\hslash}\hat{\psi}.
\label{qsh04}
\end{eqnarray}
Solving the above one can get the form of $\hat{\psi}$, namely, 
\begin{eqnarray}
\hat{\psi}(U)=C_1e^{i\frac{\mathcal{E}U}{\hslash}}.
\label{qsh05}
\end{eqnarray}
Then the wavefunction $\psi$ becomes
\begin{eqnarray}
\psi(U)=
C_1e^{i\bigg(\frac{\mathcal{E}U-\int g(U)dU}{\hslash}\bigg)}.
\label{qsh06}
\end{eqnarray}
Here
\begin{eqnarray}
\int g(U)dU=\left\{
\begin{array}{ll}
\frac{1}{2}(U\log(sec^2(\omega U))+ 2U\log(1+e^{2i\omega U})
\\ \quad +\frac{\alpha}{4} U^2-i(\omega U^2+\frac{1}{\omega}Li_2(-e^{2i\omega U}))),
&\alpha^2<4\lambda\\
\frac{(r-1)r}{\alpha (r-2)^2}e^{\frac{(r-2)}{(r-1)}U},
&\alpha^2>4\lambda\\
\frac{1}{\alpha} U^2, 
& \alpha^2=4\lambda\\
-\frac{1}{\alpha}e^{-\alpha U}, & \lambda=0
\end{array}\right.\label{qsh06b}
\end{eqnarray}
where $Li_n(z)$ is polylogarithm function \cite{Abramowitz}. In order to relate
this form of the wavefunctions in terms of the original variable one has to
make use of the appropriate canonical transformations.

\end{document}